\documentstyle[12pt]{article}
\def\mod{ {\rm mod} }

\begin{document}
\baselineskip=14pt plus 1pt minus 1pt



\begin{center} {\bf Symmetries of anisotropic harmonic oscillators with
rational ratios}

{\bf of frequencies and their relations to U(N) and O(N+1)
\footnote[1]{Presented by D. Bonatsos}} 

\end{center}

\bigskip\bigskip\bigskip

\begin{center}
Dennis Bonatsos$^{+ *}$\footnote[2]{e-mail: bonat@ectstar.ect.unitn.it,
bonat@cyclades.nrcps.ariadne-t.gr}, 
C.~Daskaloyannis$^{\dagger}$\footnote[3]{e-mail: 
daskaloyanni@olymp.ccf.auth.gr}, 
 P.~Kolokotronis$^{+}$ and D. Lenis$^{+}$\footnote[4]{e-mail: 
lenis@cyclades.nrcps.ariadne-t.gr}
\bigskip 

$^+$ Institute of Nuclear Physics, N.C.S.R. ``Demokritos''

GR-15310 Aghia Paraskevi, Attiki, Greece 

$^*$ European Centre for Theoretical Research in Nuclear Physics and Related 
Areas (ECT$^*$)

Villa Tambosi, Strada delle Tabarelle 286, I-38050 Villazzano (Trento), Italy 

$^\dagger$ Department of Physics, Aristotle University of
Thessaloniki

GR-54006 Thessaloniki, Greece 
\end{center}
\bigskip\bigskip\bigskip

\centerline{\bf Abstract}

The concept of bisection of a harmonic oscillator or hydrogen atom, used in 
the past in establishing the connection between U(3) and O(4),  is 
generalized into multisection (trisection, tetrasection, etc). It is then 
shown that all symmetries of the N-dimensional anisotropic harmonic oscillator
with rational ratios of frequencies (RHO), some of which are underlying 
the structure of superdeformed and hyperdeformed nuclei, 
can be obtained from the U(N)
symmetry of the corresponding isotropic oscillator with an appropriate 
combination of multisections. Furthermore, it is seen that bisections of 
the N-dimensional hydrogen atom, which possesses an O(N+1) symmetry, lead to 
the U(N) symmetry, so that further multisections of the hydrogen atom lead 
to the symmetries of the N-dim RHO.   
 The opposite is in general not true, i.e. multisections of U(N) do 
not lead to O(N+1) symmetries, the only exception
being the occurence of O(4) after the bisection of U(3). 

\bigskip\bigskip\bigskip
{\bf 1. Introduction}

Anisotropic harmonic oscillators with rational ratios of frequencies (RHOs)
\cite{JM,Dem,Dui,Mai,Ven,MV,Cis} are 
of current interest in several branches of physics. Their symmetries form
the basis for the understanding \cite{Mot,Rae,Ros,Bha,Naz} of the occurence of 
superdeformed and hyperdeformed nuclear shapes \cite{Nol,Jan} at very high 
angular momenta. In addition, 
 they have been recently connected \cite{RZ,ZR} to the underlying geometrical 
structure in the Bloch--Brink $\alpha$-cluster model \cite{Bri}. They are also
becoming of interest for the interpretation of the observed shell 
structure in atomic clusters \cite{Mar}, especially after the realization that 
large deformations can occur in such systems \cite{Bul}.  
An interesting problem is to what extend the various symmetries of the RHOs,
occuring for different frequency ratios, are related to other known 
symmetries. A well-known example is the case of the 3-dimensional RHO 
with frequency ratios 2:2:1, which is known to possess the O(4) symmetry
\cite{Rav}. 

In this paper we show how the symmetries of the N-dim RHO can be obtained 
from the U(N) symmetry of the corresponding isotropic harmonic oscillator 
(HO) by appropriate symmetry operations, namely multisections, which are 
generalizations of the concept of bisection, introduced in \cite{Rav}. It will 
furthermore 
be shown that these symmetries can also be obtained from the O(N+1) 
symmetry of the N-dim hydrogen atom, since a bisection leads from O(N+1)
to U(N), so that further multisections lead to RHO symmetries. 
However, despite the fact that the N-dim RHO symmetries can be 
obtained from the O(N+1) symmetry by appropriate multisections, they are
not orthogonal symmetries themselves (with the exception of 2:2:1
mentioned above). 

In Section 2 of this paper multisections of the N-dim harmonic oscillator are
defined and used in obtaining the symmetries of the various RHOs. A similar
procedure is followed in Section 3 for the N-dim hydrogen atom. Section 4
contains discussion of the present results and implications for further 
work. 

{\bf 2. Multisections of the harmonic oscillator} 

The Hamiltonian of the N-dim RHO reads  
$$ H = {1 \over 2} \sum_{k=1}^N \left( p_k^2 + {x_k^2\over m_k^2}\right),
\eqno(1)$$
where $m_i$ are natural numbers prime to each other. The energy eigenvalues
are given by 
$$ E = \sum_{k=1}^N {1\over m_k} \left( n_k+{1\over 2}\right),\eqno(2)$$
where $n_k$ is the number of quanta in the $k$-th direction. Alternatively,
the energy eigenvalues can be written as \cite{BDKL,Patras94}
$$ E =\Sigma +\sum_{k=1}^N {2 q_k-1\over 2 m_k},\eqno(3) $$
with $q_k = 1, 2, \dots, m_k$, the connection between the two pictures 
been given by 
$$ n_k = [n_k/m_k] m_k + \mod(n_k, m_k),\eqno(4) $$
$$ \Sigma= \sum_{k=1}^N [n_k/m_k], \eqno(5)$$
$$ q_k = \mod(n_k,m_k)+1,\eqno(6)$$
where $[x]$ stands for the integer part of $x$. 

{\bf 2.1 The 3-dimensional oscillator}

Let us consider the completely symmetric irreps of U(3), $[N00]$, the 
dimensions of which are given by 
$$ d(N) = {(N+1)(N+2)\over 2}, \qquad N=0, 1, 2, \ldots \eqno(7)$$  
Using the Cartesian notation  
$(n_x, n_y, n_z)$ for the U(3) states, as in \cite{Rav}, we have the following 
list:

N=0: (000)

N=1: (100) (010) (001)

N=2: (200) (020) (002) (110) (101) (011) 

N=3: (300) (030) (003) (210) (120) (201) (102) (021) (012) (111) 

N=4: (400) (040) (004) (310) (130) (301) (103) (031) (013) (220) (202) (022)
     (211) (121) (112) 

N=5: (500) (050) (005) (410) (140) (401) (104) (041) (014) (320) (230) (302)
     (203) (032) (023) (311) (131) (113) (221) (212) (122). 

We see that the corresponding degeneracies are 1, 3, 6, 10, 15, 21, \dots,  
 which
correspond to the dimensions of the U(3) irreps, as mentioned above, i.e. 
to the 1:1:1 HO.  

Choosing only the $n_z$=odd states, one is left with the following list

N=1: (001)

N=2: (101) (011)

N=3: (003) (201) (021) (111)

N=4: (301) (103) (031) (013) (211) (121)

N=5: (005) (401) (041) (203) (023) (311) (131) (113) (221),  

while 
choosing only the $n_z$=even states, one is left with the following list:

N=0: (000)

N=1: (100) (010)

N=2: (200) (020) (002) (110) 

N=3: (300) (030) (210) (120) (102) (012) 

N=4: (400) (040) (004) (310) (130) (220) (202) (022) (112) 

N=5: (500) (050) (410) (140) (104) (014) (320) (230) (302) (032) (212) (122).

We see that in both cases the degeneracies are  1, 2, 4, 6, 9, 12, 16, 20, 
\dots, which are the degeneracies of the 3-dim RHO with ratios 2:2:1. 
Therefore a {\bf bisection} of the 1:1:1 RHO states, distinguishing 
states with $\mod(n_z,2)=0$ and states with $\mod(n_z,2)=1$,  results it two
interleaving 2:2:1 sets of levels. 

By analogy, a {\bf trisection} can be made  by distinguishing states 
with $\mod(n_z,3)=0$ or $\mod(n_z,3)=1$ or $\mod(n_z,3)=2$. For 
$\mod(n_z,3)=0$ we obtain 

N=0: (000)

N=1: (100) (010)

N=2: (200) (020) (110) 

N=3: (300) (030) (003) (210) (120)

N=4: (400) (040) (310) (130) (103) (013) (220) 

N=5: (500) (050) (410) (140) (320) (230) (203) (023) (113),  

while for $\mod(n_z,3)=1$ one has 

N=1: (001) 

N=2: (101) (011) 

N=3: (201) (021) (111)

N=4: (004) (301) (031) (211) (121)

N=5: (401) (104) (041) (014) (311) (131) (221), 

and for $\mod(n_z,3)=2$ one has 

N=2: (002)

N=3: (102) (012) 

N=4: (202) (022) (112)

N=5: (005) (302) (032) (212) (122). 

The degeneracies obtained are 1, 2, 3, 5, 7, 9, 12, 15, 18, \dots, which 
correspond 
to the 3:3:1 RHO. Therefore a trisection of the 1:1:1  HO results in three
interleaving sets of 3:3:1 RHO states. 

Similarly a {\bf tetrasection} is defined by selecting states with 
$\mod(n_z,4)=0$,  or 1, or 2, or 3. In the case of $\mod(n_z,4)=0$ one has

N=0: (000)

N=1: (100) (010)

N=2: (200) (020) (110)

N=3: (300) (030) (210) (120)

N=4: (400) (040) (004) (310) (130) (220)

N=5: (500) (050) (410) (140) (104) (014) (320) (230),   

while for $\mod(n_z,4)=1$ one obtains

N=1: (001)

N=2: (101) (011)

N=3: (201) (021) (111)

N=4: (301) (031) (211) (121)

N=5: (005) (401) (041) (311) (131) (221), 

for $\mod(n_z,4)=2$ one has 

N=2: (002)

N=3: (102) (012)

N=4: (202) (022) (112)

N=5: (302) (032) (212) (122),

and for $\mod(n_z,4)=3$ one gets 

N=3: (003) 

N=4: (103) (013)

N=5: (203) (023) (113) . 

The degeneracies obtained are 1, 2, 3, 4, 6, 8, 10, 12, 15, 18, \dots,
which characterize the 4:4:1 RHO.   Therefore a tetrasection of the 
1:1:1 HO leads to four interleaving sets of 4:4:1 RHO states. 

In general, an {\sl n-section} of the 1:1:1 HO is obtained by separating 
states with $mod(n_z, n)$ =0, or 1, or 2, \dots, or $n-1$. In this case 
$n$ interleaving sets of the n:n:1 RHO states, which corresponds to an 
oblate shape,  are obtained. 
It is clear that n-sections using $n_x$ or $n_y$ instead of $n_z$ lead 
to the same conclusions. 

One can consider successively more than one bisections, trisections, etc. 
Let us consider more than one bisections first. 

Getting the results of the $\mod(n_z,2)=0$ bisection of the HO and 
applying a $\mod(n_y,2)$ =0 bisection on them we obtain 

N=0: (000)

N=1: (100)

N=2: (200) (020) (002) 

N=3: (300) (120) (102) 

N=4: (400) (040) (004) (220) (202) (022) 

N=5: (500) (140) (104) (320) (302) (122). 

The degeneracy pattern is 1, 1, 3, 3, 6, 6, 10, 10, i.e. ``two copies''
of the 1:1:1 degeneracies,  which corresponds 
to the 2:1:1 RHO. The same result is obtained for any combination of 
two bisections along two differerent axes. 

Bisecting the 1:1:1 HO for a third time, along the $x$-axis this time by
using $\mod(n_x,2)$ =0, one obtains

N=0: (000) 

N=2: (200) (020) (002)

N=4: (400) (040) (004) (220) (202) (022). 

The degeneracy pattern is 1, 3, 6, 10, \dots, i.e. that of the original
1:1:1 HO. 

Furthermore one can easily see that:

i) Two trisections along different axes lead to degeneracies 1, 1, 1, 3, 3, 3, 
6, 6, 6, \dots, i.e. to the 3:1:1 RHO pattern (``three copies'' of the 
1:1:1 degeneracies).  

ii) Three trisections lead to the original 1:1:1 HO degeneracy pattern. 

iii) Two tetrasections lead to degeneracies 1, 1, 1, 1, 3, 3, 3, 3, 6, 6, 6, 6,
\dots, i.e. to the 4:1:1 RHO pattern (``four copies'' of the 1:1:1 
degeneracies).  

iv) Three tetrasections lead back to the original 1:1:1 HO pattern. 

The results obtained so far are summarized in Table 1. 

\begin{table}
\centerline{\bf Table 1}

Degeneracies of various 3-dim anisotropic harmonic oscillators with 
rational ratios of frequencies (RHOs)  obtained from the U(3) symmetry of the 
isotropic 3-dim
harmonic oscillator (HO) by the application of various multisections. 
The first line corresponds to the isotropic 3-dim HO. In the rest of 
the lines the first column contains the appropriate multisection, while the 
second column contains the frequency ratios $m_1:m_2:m_3$ of the resulting 
RHO. \bigskip
\hrule
$$\vbox{\halign{\hfil #\hfil &&\quad \hfil #\hfil \cr
 U(3)  &  1:1:1 & 1 & 3 & 6 & 10 & 15 & 21 & 28 & 36 &  & & & & \cr
1 bisection &  
2:2:1 & 1 & 2 & 4 & 6 & 9 & 12 & 16 & 20 & & & & & \cr
1 trisection &
3:3:1 & 1 & 2 & 3 & 5 & 7 & 9 & 12 & 15 & 18 & 22 & 26 & 30 &  \cr
1 tetrasection & 
4:4:1 & 1 & 2 & 3 & 4 & 6 & 8 & 10 & 12 & 15 & 18 & 21 & 24 & 28 \cr 
2 bisections & 
1:1:2 & 1 & 1 & 3 & 3 & 6 & 6 & 10 & 10 & 15 & 15 & 21 & 21 & 28 \cr
2 trisections & 
1:1:3 & 1 & 1 & 1 & 3 & 3 & 3 & 6 & 6 & 6 & 10 & 10 & 10 & 15 \cr
2 tetrasections & 
1:1:4 & 1 & 1 & 1 & 1 & 3 & 3 & 3 & 3 & 6 & 6 & 6 & 6 & 10 \cr
3 bisections & 1:1:1 & 1 & 3 & 6 & 10 & 15 & 21 & 28 & 36 & & & & & \cr
3 trisections & 1:1:1 & 1 & 3 & 6 & 10 & 15 & 21 & 28 & 36 & & & & & \cr
3 tetrasection & 1:1:1 & 1 & 3 & 6 & 10 & 15 & 21 & 28 & 36 & & & & & \cr
}}$$
\hrule
\end{table}
In general one can see that: 

i) Two n-sections (along different axes) lead to the degeneracy pattern of 
n:1:1, i.e. to ``n copies'' of the 1:1:1 degeneracies. n:1:1 corresponds 
to a prolate shape.  

ii) Three n-sections lead back to the degeneracy pattern of the 1:1:1 HO. 

One can, of course, apply successive n-sections with different n. For
example, applying $\mod(n_z,2)=0$, $\mod(n_y,3)=0$ and $\mod(n_x,3)=0$
one obtains the degeneracy pattern 1, 1, 2, 1, 2, 4, 2, 4, 6, \dots, 
which corresponds to the 2:2:3 oscillator.  

In general one can see that by applying a $k$-section, an $l$-section and 
an $m$-section along different axes one obtains the degeneracy pattern 
$(kl):(mk):(lm)$, where common factors appearing in all three quantities 
$(kl)$, $(mk)$, $(lm)$ can be dropped out.  

We have therefore seen that all the symmetries of the 3-dim RHO can be 
obtained 
from the U(3) symmetry of the isotropic 3-dim HO by an appropriate set of 
n-sections. 

A special remark can be made about the 2:2:1 case. The degeneracies obtained 
there correspond to the dimensions of the irreps of O(4), given by 
$$ d(\mu_1, \mu_2) = (\mu_1+\mu_2+1) (\mu_1-\mu_2+1).\eqno(8)$$ 
In particular, the degeneracies 1, 4, 9, 16, \dots correspond to the 
integer irreps $(\mu,0)$ with $\mu=0$, 1, 2, 3, \dots, while the degeneracies 
2, 6, 12, 20, \dots correspond to the spinor irreps $({n\over 2}, {1\over 2})$
with $n=1$, 3, 5, 7, \dots. This result has been first found by Ravenhall
{\it et al.} \cite{Rav}. It has been pointed out that O(4) is obtained by 
imposing a reflection condition on U(3). For example, O(4) is obtained by 
 selecting the states with $n_z$=odd, a procedure which is equivalent to 
the insertion of an impenetrable barrier across the $xy$ plane. 

{\bf 2.2 The 4-dimensional oscillator} 

\begin{table}
\centerline{\bf Table 2}

Same as Table 1, but for the  4-dim oscillator.
\bigskip \hrule
$$\vbox{\halign{\hfil #\hfil &&\quad \hfil #\hfil \cr
   & &   &   &    &    &    &    &    &    &    &    \cr
U(4) & 1:1:1:1  & 1 & 4 & 10 & 20 & 35 & 56 &  &    &    &    \cr
1 bisection & 2:2:2:1 & 1 & 3 & 7 & 13  &  22 &  &  &  &  &  \cr
2 bisections & 2:2:1:1 & 1 & 2 & 5 &  8 &  14 &  20 & 30 & 40 & 55 & 70 \cr 
3 bisections & 2:1:1:1 & 1 & 1 & 4 &  4 & 10 & 10 &    &    &    &    \cr
4 bisections & 1:1:1:1 & 1 & 4 & 10 & 20 & 35 & 56 & & & & \cr
1 trisection & 3:3:3:1 & 1 & 3 & 6 & 11 & 18 & 27 & 35 &    &    &    \cr
2 trisections & 3:3:1:1 & 1 & 2 & 3 & 6 &  9 &    &    &    &    &    \cr
3 trisections & 3:1:1:1 & 1 & 1 & 1 & 4 &  4 & 4  & 10 & 10 & 10 & 20 \cr
4 trisections & 1:1:1:1 & 1 & 4 & 10 & 20 & 35 & 56 & & & & \cr 
1 tetrasection & 4:4:4:1 & 1 & 3 & 6 & 10 &  &    &    &    &    &    \cr 
}}$$
\hrule
\end{table}

The relevant information is given in Table 2. 
The symmetry of the HO in this case is U(4). The first line of the table 
corresponds to the dimensions of the symmetric irreps of U(4), 
$[ N, 0, 0, 0]$, given by the equation
$$ d(N) = {1\over 6} (N+1)(N+2)(N+3) .\eqno(9)$$
Notice that these degeneracies coincide with the dimensions of the 
symmetric irreps $(N, 0)$ of Sp(4). (It is known that Sp(4) is a
subalgerbra of U(4).) 

One bisection of U(4) leads to the 2:2:2:1 degeneracies.

Two bisections of U(4) lead to the 2:2:1:1 degeneracies
 1, 2, 5, 8, 14, 20, 30, 40, 55, 70, \dots. 
Out of these, 1, 5, 14, 30, 55, \dots correspond to the symmetric irreps   
$(\mu, 0)$ of O(5), while 2, 8, 20, 40, 70, \dots  correspond to the 
$({n\over 2}, {1\over 2})$ irreps of O(5). The relevant
formula is:
$$ d(m_1, m_2) = {1\over 6} (2 m_1+3) (2 m_2+1) (m_1+m_2+2) (m_1-m_2+1).
\eqno(10) $$
Notice that the dimensions of the integer irreps (1, 5, 14, 30, 55, \dots) are 
reproduced exactly, while the dimensions of the half-integer irreps 
are half of the ones given by eq. (10), which are 4, 16, 40, 80, 140, \dots, 
respectively. Therefore the 2:2:1:1 symmetry is not O(5), although it bears 
certain similarities to it. 

The occurence of a  symmetry resembling O(5) is not surprising, since 
 U(4) is isomorphic to O(6), which does have an O(5) subalgebra. 
The generators of the O(5) subalgebra in terms of the U(4) generators have 
been given explicitly in \cite{Hec,Hol}. It is also known that O(5) is 
isomorphic 
to Sp(4), which is a subalgebra of U(4), since in general U(2n) possesses an
Sp(2n) subalgebra. 

Three bisections of U(4) lead to the 2:1:1:1 degeneracies, i.e. to ``two 
copies'' of the U(4) degeneracies. 

Finally, four bisections of U(4) lead back to the U(4) degeneracies 
characterizing the 1:1:1:1 HO. 

Similarly one can see that a trisection of U(4) leads to the 3:3:3:1 
degeneracies, two trisections of U(4) lead to 3:3:1:1, three trisections 
of U(4) lead to 3:1:1:1, i.e. to ``three copies''  of the U(4) 
degeneracies, while four trisections of U(4) lead back to U(4). 

{\bf 2.3 The 5-dimensional  oscillator}

\begin{table}
\centerline{\bf Table 3}

Same as Table 1, but for the 5-dim oscillator.
\bigskip\hrule
$$\vbox{\halign{\hfil #\hfil &&\quad \hfil #\hfil \cr
U(5) & 1:1:1:1:1 & 1 & 5 & 15 & 35 & 70 &  &  &  &    &    &    \cr
1 bisection & 2:2:2:2:1 & 1 & 4 & 11 & 24  &  46 &  &  &  &  &  & \cr
2 bisections & 2:2:2:1:1 & 1 & 3 & 8 &  16 &  30 &   &  &  &  &  &    \cr 
3 bisections & 2:2:1:1:1  
& 1 & 2 & 6 & 10 & 20 & 30 & 50 & 70 & 105 & 140 & 196   \cr
4 bisections & 2:1:1:1:1 
& 1 & 1 & 5 & 5  & 15 & 15 &    &    &     &     &       \cr
5 bisections & 1:1:1:1:1 & 1 & 5 & 15 & 35 & 70 & & & & & & \cr 
}}$$
\hrule
\end{table}

The relevant results are shown in Table 3. In the first line the dimensions
of the symmetric irreps $[N, 0, 0, 0, 0]$ of U(5) appear, given by
the formula
$$ d(N)= {1\over 24} (N+1) (N+2) (N+3) (N+4) .\eqno(11) $$
 There is no
symplectic subalgebra in this case. 

One bisection of U(5) leads to the 2:2:2:2:1 degeneracies, while two
bisections lead to the 2:2:2:1:1 pattern. 

Three bisections lead to the 2:2:1:1:1 degeneracies 1, 2, 6, 10, 20, 30, 
50, 70, 105, 140, 196, \dots.  In particular, the degeneracies 1, 6, 20, 50, 
105, 196, \dots correspond to the dimensions of the integer irreps 
$(N, 0, 0)$ of O(6), while the intermediate degeneracies 2, 10, 30, 70, 140, 
\dots  resemble  the half-integer
irreps $({n\over 2}, {1\over 2}, {1\over 2})$ of O(6). 
The relevant formula is
$$ d(m_1, m_2, m_3) = {1\over 12} (m_1+m_2+3) (m_1+m_3+2) (m_2+m_3+1)$$
$$(m_1-m_2+1) (m_1-m_3+2) (m_2-m_3+1) .\eqno(12) $$
Again there is a factor of 2 difference for the half-integer irreps: The 
results in Table 3 are ${1\over 2}$ times the results given by the above
equation. Therefore the 2:2:1:1:1 symmetry is not an O(6) symmetry. 

Finally, four bisections lead to the 2:1:1:1:1 degeneracy pattern, while 
five bisections lead back to the U(5) degeneracies. 

{\bf 2.4 The 6-dimensional oscillator}

\begin{table}
\centerline{\bf Table 4}

Same as Table 1, but for the 6-dim oscillator.
\bigskip
\hrule
$$\vbox{\halign{\hfil #\hfil &&\quad \hfil #\hfil \cr
U(6) & 1:1:1:1:1:1 & 1 & 6 & 21 & 56 &  &  & &    &    &    \cr
1 bisection & 2:2:2:2:2:1 & 1 & 5 & 16 &   &   &  &  & &  & \cr
2 bisections & 2:2:2:2:1:1 & 1 & 4 & 12 &   &   &   &  &  &  &    \cr 
3 bisections & 2:2:2:1:1:1 & 1 & 3 & 9 &  &  &  &  &    &    &    \cr
4 bisections & 2:2:1:1:1:1 
& 1 & 2 & 7 & 12 & 27 & 42 & 77 & 112 & 182 & 252 \cr
5 bisections & 2:1:1:1:1:1 
& 1 & 1 & 6 & 6 &  21 & 21 &    &     &     &     \cr
6 bisections & 1:1:1:1:1:1 & 1 & 6 & 21 & 56 & & & & & & \cr
}}$$
\hrule
\end{table}

In this case the symmetry is U(6). The relevant results are given in Table 4.
In the first line, the dimensions of the symmetric irreps $[N,0,0,0,0,0]$
of U(6) appear, given by the formula
$$ d(N) = {1\over 120} (N+1)(N+2)(N+3)(N+4)(N+5).\eqno(13) $$
 They coincide with the dimensions of the symmetric irreps 
$(N,0,0)$ of Sp(6). It is known that Sp(6) is a subalgebra of U(6). 

One bisection leads to the 2:2:2:2:2:1 degeneracy pattern, while two 
bisections lead to 2:2:2:2:1:1 and three bisections lead to 2:2:2:1:1:1. 

Four  bisections lead to degeneracies resembling the dimensions of O(7) 
irreps. The degeneracies  1, 7, 27, 77, 182, \dots  
correspond to the dimensions of integer O(7) irreps of the form $(m, 0, 0)$, 
while the numbers 2, 12, 42, 112, 252, \dots resemble the dimensions of the 
half-integer irreps
$({m\over 2}, {1\over 2}, {1\over 2})$ of O(7). The relevant formula is
$$ d(m_1,m_2,m_3)= {1\over 720} (2 m_1+5) (2 m_2 +3) (2 m_3+1)$$ $$(m_1+m_2+4) 
(m_1+m_3+3) (m_2+m_3+2)$$ $$ (m_1-m_2+1) (m_1-m_3+2) (m_2-m_3+1).\eqno(14)$$
This formula gives for the integer irreps the results of Table 4, but for 
the half-integer irreps it gives 4 times the results of Table 4. Therefore 
the 2:2:1:1:1:1 symmetry is not an O(7) symmetry. 

{\bf 2.5 The N-dimensional oscillator}

The symmetry is U(N). If N is even, there is an Sp(N) subalgebra, if N is odd
there is no such subalgebra. 

N bisections lead back to the U(N) irreps. 

N-1 bisections lead to the 2:1:1:\dots:1 symmetry, i.e. to ``two copies''
 of the U(N) irreps. 

N-2 bisections lead to the 2:2:1:1:\dots :1 symmetry, which bears certain
similarities to O(N+1). The dimensions of the integer irreps 
are obtained correctly. The dimensions of the odd irreps differ 
by a factor of $2^{\nu-1}$, where $\nu=N/2$ for $N$ even or $\nu=(N-1)/2$ for 
$N$ odd. Therefore the 2:2:1:1:\dots :1 symmetry is not in general O(N+1).   

N-3 bisections lead to the 2:2:2:1:1:\dots :1 degeneracies. 

Two bisections lead to the 2:2:\dots :2:1:1 degeneracies. 

One bisection leads to the 2:2:\dots :2:2:1 degeneracies. 

Similarly

one n-section leads to the n:n:\dots :n:n:1 degeneracies, 

two n-sections lead to the n:n:\dots :n:1:1 degeneracies, 

N-2 n-sections lead to n:n:1:\dots :1:1, 

N-1 n-sections lead to n:1:1:\dots :1:1, 

N n-sections lead back to U(N). 

{\bf 3. Multisections of the hydrogen atom} 

So far we have considered multisections of the N-dim harmonic 
oscillator. We are now going to consider multisections of the hydrogen 
atom (HA) in N dimensions, which is known to be characterized by the O(N+1)
symmetry \cite{All}, which is also the symmetry characterizing a particle 
constrained to move on an (N+1)-dim hypersphere. 

{\bf 3.1 The 3-dimensional hydrogen atom}

The 3-dim hydrogen atom is known to possess the O(4) symmetry. 
We know that the irreps of O(4) are characterized by two labels $\mu_1$, 
$\mu_2$ and are denoted by $(\mu_1, \mu_2)$, while the irreps of O(3) are 
characterized by one label $\mu_1'=L $ (the usual angular momentum 
quantum number) and are denoted  by $(\mu_1')$. When making the reduction 
O(4) $\supset$ O(3), $\mu_1'$ obtains all values permitted by the condition 
$\mu_1 \geq \mu_1' \geq \mu_2$ \cite{Ham}. Furthermore, the decomposition 
O(3)$\supset$O(2) can be made, the irreps of O(2) characterized  by the 
quantum number $M=L$, $L-1$, $L-2$, \dots, $-(L-1)$, $-L$. 

\begin{table}
\centerline{\bf Table 5}

Decomposition of completely symmetric O(4) irreps (corresponding to the 
3-dim hydrogen atom) using the O(4) $\supset$ O(3) $\supset$ O(2) chain. 
In the last column the states are labelled by $(LM)$, where $L$ is the 
O(3) quantum number and $M$ is the O(2) one.   
\bigskip\hrule
$$\vbox{\halign{\hfil #\hfil &&\quad \hfil #\hfil \cr 
O(4)       &   O(3)       &    (LM)     \cr
(00) & (0)       &   (00) \cr
(10) & (1) (0)   &   (11) (10) (1-1) (00) \cr
(20) & (2) (1) (0) & (22) (21) (20) (2-1) (2-2) (11) (10) (1-1) (00) \cr 
(30) & (3) (2) (1) (0) & (33) (32) (31) (30) (3-1) (3-2) (3-3) \cr
     &                 &   (22) (21) (20) (2-1) (2-2) (11) (10) (1-1) (00) \cr
(40) & (4) (3) (2) (1) (0) & (44) (43) (42) (41) (40) (4-1) (4-2) (4-3) (4-4)
\cr
     &                  & (33) (32) (31) (30) (3-1) (3-2) (3-3) \cr
     &                  & (22) (21) (20) (2-1) (2-2) (11) (10) (1-1) (00) \cr
(50) & (5) (4) (3) (2) (1) (0) & (55) (54) (53) (52) (51) (50) (5-1) (5-2) 
(5-3) (5-4) (5-5) \cr
     &                 & (44) (43) (42) (41) (40) (4-1) (4-2) (4-3) (4-4) \cr
     &                 & (33) (32) (31) (30) (3-1) (3-2) (3-3) \cr
     &                 & (22) (21) (20) (2-1) (2-2) (11) (10) (1-1) (00) \cr
}}$$
\hrule
\end{table}

We are going to consider the completely symmetric irreps of O(4), which are 
of the form $(\mu_1, 0)$. The $(LM)$ states contained in each O(4) irrep 
are shown in Table 5. The dimensions of the irreps are 1, 4, 9, 16, 25, \dots,
as expected from eq. (8), since only the integer irreps occur. 
As pointed out by Ravenhall {\it et al.} \cite{Rav}, a bisection
can be effected by inserting an impenetrable barrier through the center 
of the hydrogen atom. Only the states with $L-M$=odd remain then. 
From Table 5 one sees that the remaining states are:

$\mu_1=1$: (10) 

$\mu_1=2$: (10) (21) (2-1) 

$\mu_1=3$: (10) (21) (2-1) (32) (30) (3-2) 

$\mu_1=4$: (10) (21) (2-1) (32) (30) (3-2) (43) (41) (4-1) (4-3)

$\mu_1=5$: (10) (21) (2-1) (32) (30) (3-2) (43) (41) (4-1) (4-3) 
           (54) (52) (50) (5-2) (5-4), 

which correspond to degeneracies 1, 3, 6, 10, 15, \dots, i.e. the degeneracies 
of U(3). 

Keeping the states with $L-M$=even one is left with 

$\mu_1=0$: (00)

$\mu_1=1$: (00) (11) (1-1) 

$\mu_2=2$: (00) (11) (1-1) (22) (20) (2-2) 

$\mu_1=3$: (00) (11) (1-1) (22) (20) (2-2) (33) (31) (3-1) (3-3) 

$\mu_1=4$: (00) (11) (1-1) (22) (20) (2-2) (33) (31) (3-1) (3-3) 
           (44) (42) (40) (4-2) (4-4) 

$\mu_1=5$: (00) (11) (1-1) (22) (20) (2-2) (33) (31) (3-1) (3-3) 
           (44) (42) (40) (4-2) (4-4) (55) (53) (51) (5-1) (5-3) (5-5). 

The resulting degeneracies are again 1, 3, 6, 10, 15, 21, \dots, i.e. 
U(3) degeneracies. Therefore a {\bf bisection} of the 3-dim hydrogen atom, 
effected by choosing states with $\mod(L-M, 2)=0$ or $\mod(L-M,2)=1$,  
is leading to two interleaving sets of U(3) states. Choosing states 
with $\mod(L+M, 2)=0$ or 1 obviously leads to the same results. 

The fact that by bisecting O(4) one obtains U(3) has been first pointed
out by Ravenhall {\it et al.} \cite{Rav}. 
Once the U(3) symmetry of the 3-dim HO
is obtained, any further multisections on it will lead to RHO degeneracies,
as pointed out in subsec. 2.1.  We briefly show how this can be carried out by 
a few examples. 

i) Selecting states with $\mod(L-M, 4)=0$ gives 

$\mu_1=0$: (00)

$\mu_1=1$: (00) (11) 

$\mu_1=2$: (00) (11) (22) (2-2) 

$\mu_1=3$: (00) (11) (22) (2-2) (33) (3-1) 

$\mu_1=4$: (00) (11) (22) (2-2) (33) (3-1) (44) (40) (4-4),  

\noindent
i.e. it leads to degeneracies 1, 2, 4, 6, 9, \dots, which are those 
of the 2:2:1 RHO. The same result is obtained by choosing states with
$\mod(L-M,4)=2$. Therefore the operation of dividing the states with 
$\mod(L-M,2)=0$
of the 3-dim HA  
according to the $\mod(L-M,4)$ is equivalent to a bisection of the 3-dim HO. 
The same holds for $\mod(L+M,4)$, as well as for dividing the $\mod(L-M,2)=1$
states of the 3-dim HA according to $\mod(L-M,4)=1$ or 3.   

ii) Selecting states with $\mod(L-M,6)=0$, or 2, or 4 (or 1, or 3, or 5) leads
to degeneracies 1, 2, 3, 5, 7, 9, 12, \dots, i.e. to the degeneracies 
of the 3:3:1 RHO. Therefore this operation is equivalent to a trisection 
of the 3-dim HO. The same holds for $\mod(L+M,6)$. 

iii) Selecting states with $\mod(L-M,8)=0$, or 2, or 4, or 6 (or 1, or 3, or 5, or 7)
leads to degeneracies 1, 2, 3, 4, 6, 8, 10, 12, \dots, i.e. to the 
degeneracies of the 4:4:1 RHO. Therefore this operation is equivalent to a
tetrasection of the 3-dim HO. The same holds for $\mod(L+M,8)$. 

Combining two of the above operations one obtains the results corresponding
to the appropriate multisections of the HO. Thus:

i) Selecting states with $\mod(L-M,4)=0$ and $\mod(L+M,4)=0$ one finds the
degeneracies 1, 1, 3, 3, 6, 6, 10, 10, \dots, which correspond to the 
2:1:1 RHO. 

ii) Selecting states with $\mod(L-M,6)=0$ and $\mod(L+M,6)=0$ one finds the
degeneracies 1, 1, 1, 3, 3, 3, 6, 6, 6, \dots, which characterize the
3:1:1 RHO. 

iii) Selecting states with $\mod(L-M,8)=0$ and $\mod(L+M,8)=0$ one finds the 
degeneracies 1, 1, 1, 1, 3, 3, 3, 3, 6, 6, 6, 6, \dots, which 
correspond to the 4:1:1 RHO. 

{\bf 3.2 The 4-dimensional hydrogen atom} 

The 4-dim hydrogen atom is characterized by the O(5) symmetry. The irreps 
of O(5) can be  labelled as $(\mu_1, \mu_2)$, while the irreps of  O(4)
can be labelled by $(\mu_1', \mu_2')$. When making the reduction 
O(5)$\supset$O(4), $\mu_1'$ and $\mu_2'$ take the values permitted by the 
relation $\mu_1 \geq \mu_1' \geq \mu_2 \geq \mu_2'$ \cite{Ham}.
 Continueing further 
the reduction O(5)$\supset$O(4)$\supset$O(3)$\supset$O(2) one obtains 
the lists of states given it Table 6.   The dimensions of the irreps are
1, 5, 14, 30, 55, \dots, as expected from eq. (10), since only the 
integer irreps occur. 

\begin{table}
\centerline{\bf Table 6} 

Decomposition of completely symmetric O(5) irreps (corresponding to the 
4-dim hydrogen atom) using the O(5) $\supset$ O(4) $\supset$ O(3) chain. 
\bigskip\hrule
$$\vbox{\halign{\hfil #\hfil &&\quad \hfil #\hfil \cr
O(5)      &     O(4)      &    O(3)     \cr
(00)      & (00)          &  (0)        \cr
(10)      & (10) (00)     & (1) $(0)^2$  \cr
(20)      & (20) (10) (00) & (2) $(1)^2$ $(0)^3$ \cr
(30)      & (30) (20) (10) (00) & (3) $(2)^2$  $(1)^3$ $(0)^4$ \cr 
(40)      & (40) (30) (20) (10) (00) & (4) $(3)^2$ $(2)^3$ $(1)^4$ $(0)^5$
 \cr}}$$
\hrule
\end{table}

Selecting states with $\mod(L-M,2)=0$ or 1 one obtains the degeneracies 
1, 4, 10, 20, 35, \dots, which characterize U(4), i.e. the 4-dim 
isotropic HO 1:1:1:1. 

Bisecting these results using $\mod(L-M,4)$ one obtains the degeneracies 
1, 3, 7, 13, 22, \dots, which correspond to the 2:2:2:1 RHO, while 
trisecting them according to $\mod(L-M,6)$ one obtains the degeneracies 
1, 3, 6, 11, 18, \dots, which correspond to the 3:3:3:1 RHO, and 
tetrasecting them
according to $\mod(L-M,8)$ one obtains the degeneracies 1, 3, 6, 10, \dots, 
which are the degeneracies of the 4:4:4:1 RHO. 

Combining the bisections $\mod(L-M,4)$ and $\mod(L+M,4)$ one obtains 
the degeneracies 1, 2, 5, 8, \dots of the 2:2:1:1 RHO, while combining
of the trisections $\mod(L-M,6)$ and $\mod(L+M,6)$ leads to the 1, 2, 3, 6, 9,
\dots degeneracies of the 3:3:1:1 RHO. 

The similarity between the 2:2:1:1 degeneracies and the O(5) degeneracies 
can now be understood as due to the fact that the 2:2:1:1 degeneracies 
are obtained from the O(5) ones using the appropriate series of bisections
described above. 

{\bf 3.3 The 5-dimensional  hydrogen atom} 

The 5-dim hydrogen atom is characterized by the O(6) symmetry, the irreps
of which can be labelled as $(\mu_1,\mu_2,\mu_3)$, while the irreps of 
O(5) can be labelled as $(\mu_1', \mu_2')$. In the reduction 
O(6)$\supset$O(5) the labels $\mu_1'$ and $\mu_2'$ have to satisfy the 
conditions $\mu_1\geq \mu_1'\geq \mu_2 \geq \mu_2' \geq \mu_3$ \cite{Ham}.  
Continueing further 
the reduction O(6)$\supset$O(5)$\supset$O(4)$\supset$O(3)$\supset$O(2) 
one obtains the results of Table 7. The dimensions of the irreps are 
1, 6, 20, 50, \dots, as expected from eq. (12), since only integer irreps
occur. 

\begin{table}
\centerline {\bf Table 7} 

Decomposition of completely symmetric O(6) irreps (corresponding to the 
5-dim hydrogen atom) using the O(6) $\supset$ O(5) $\supset$ O(4) $\supset$
O(3) chain. 
\bigskip\hrule
$$\vbox{\halign{\hfil #\hfil &&\quad \hfil #\hfil \cr
O(6)    &   O(5)     &   O(4)   &    O(3)  \cr
(000)   & (00)    &  (00)   & (0)       \cr
(100)   & (10) (00) & (10) $(00)^2$ & (1) $(0)^3$ \cr
(200)   & (20) (10) (00) & (20) $(10)^2$ $(00)^3$ & (2) $(1)^3$ $(0)^6$ \cr 
(300)  & (30) (20) (10) (00) & (30) $(20)^2$ $(10)^3$ $(00)^4$ & (3) $(2)^3$ 
$(1)^6$ $(0)^{10}$ \cr}}$$
\hrule
\end{table}

Selecting states with $\mod(L-M,2)=0$ or 1 one obtains the degeneracies 
1, 5, 15, 35, \dots which characterize U(5), i.e. the isotropic 5-dim HO
1:1:1:1:1.

A further bisection using $\mod(L-M,4)$ leads to the  degeneracies of the 
2:2:2:2:1 RHO, while an additional bisection of these results using 
$\mod(L+M,4)$ leads to the degeneracies of the 2:2:2:1:1 RHO. 

{\bf 3.4 The 6-dimensional hydrogen atom} 

The 6-dim hydrogen atom is characterized by the O(7) symmetry, the irreps 
of which are labelled by $(\mu_1,\mu_2,\mu_3)$, while the irreps of O(6)
can be labelled by $(\mu_1',\mu_2',\mu_3')$. In the reduction 
O(7)$\supset$O(6) the labels $\mu_1'$, $\mu_2'$, $\mu_3'$ have to satisfy
the condition $\mu_1\geq \mu_1'\geq \mu_2 \geq \mu_2'\geq \mu_3 \geq \mu_3'$
\cite{Ham}. 
Further continueing the reduction 
O(7)$\supset$O(6)$\supset$O(5)$\supset$O(4)$\supset$O(3)$\supset$O(2)
one obtains the results of Table 8. The dimensions of the irreps are 1, 7, 
27, 77, \dots, as expected from eq. (14), since only integer irreps occur. 

\begin{table}
\centerline{\bf Table 8}

Decomposition of completely symmetric O(7) irreps (corresponding to the
6-dim hydrogen atom) using the O(7) $\supset$ O(6) $\supset$ O(5) $\supset$
O(4) $\supset$ O(3) chain.  
\bigskip\hrule
$$\vbox{\halign{\hfil #\hfil &&\quad \hfil #\hfil \cr 
O(7)    &   O(6)   & O(5)   &  O(4)   &   O(3)   \cr
(000)   & (000)  & (00) & (00)    &  (0)     \cr
(100)   & (100) (000) & (10) $(00)^2$ & (10) $(00)^3$ &(1) $(0)^4$ \cr
(200)   & (200) (100) (000)& (20) $(10)^2$ $(00)^3$ & 
(20) $(10)^3$ $(00)^6$ & (2) $(1)^4$ $(0)^{10}$ \cr
(300) & (300) (200) (100) (000) & (30) $(20)^2$ $(10)^3$ $(00)^4$ &
(30) $(20)^3$ $(10)^6$ $(00)^{10}$ & (3) $(2)^4$ $(1)^{10}$ $(0)^{20}$ \cr}}$$
\hrule
\end{table}

Selecting states with $\mod(L-M,2)=0$ or 1 one obtains the degeneracies 
1, 6, 21, 56, \dots which characterize U(6), i.e. the isotropic 6-dim HO
1:1:1:1:1:1. 

A further bisection using $\mod(L-M,4)$ leads to the degeneracies of the 
2:2:2:2:2:1 RHO, while an additional bisection of these results using 
$\mod(L+M,4)$ leads to the degeneracies of the 2:2:2:2:1:1 RHO. 

{\bf 3.5 The N-dimensional hydrogen atom}

The N-dim hydrogen atom is characterized by the O(N+1) symmetry. Only the 
completely symmetric irreps of O(N+1) occur. Using the chain 
O(N+1) $\supset$ O(N) $\supset$ \dots $\supset$ O(3) $\supset$ O(2) one can 
find the (LM) states contained in each O(N+1) irrep. Bisecting them using 
$\mod(L-M,2)=0$ or 1 one is left with the irreps of U(N). Further multisections
of the U(N) irreps lead to the appropriate symmetries of the N-dim RHO. It is
therefore clear that all  symmetries of the N-dim RHO can be 
obtained from a common parent, the O(N+1) symmetry. Thus it is not 
surprising that some of them (notably the 2:2:1:\dots:1 ones) show 
similarities to the corresponding O(N+1) symmetry. However, the only case 
in which an N-dim  RHO symmetry is identical to an O(N+1) symmetry occurs 
for N=3, for which the 2:2:1 RHO symmetry is O(4) \cite{Rav}. The rest of the 
RHO symmetries are not related to any orthogonal symmetries. 

{\bf 4. Discussion}

The concept of bisection of an N-dim isotropic harmonic oscillator with U(N) 
symmetry, introduced by Ravenhall {\it et al.} \cite{Rav}, has been 
generalized. Trisections, tetrasections, \dots,
n-sections of the N-dim isotropic harmonic oscillator have been introduced. 
They are shown to lead to the various symmetries of the anisotropic 
N-dim harmonic oscillator with rational ratios of frequencies (RHO).
Furthermore, multisections of the N-dim hydrogen atom with O(N+1) symmetry
have been considered. It is shown that a bisection of O(N+1) leads to U(N),
so that further multisections just lead to various cases of the N-dim RHO.   
The opposite does not hold, i.e. multisections of U(N) do not lead to
O(N+1) symmetries, the only exception being the bisection of U(3) which
does lead to O(4). Even in the case of the 4-dim HO, which has the U(4) 
symmetry, which is isomorphic to O(6) and has an O(5) subalgebra, no 
multisection, or combination of multisections, leading to a RHO with O(5)
symmetry has  been  found. We conclude therefore that the rich variety of
the N-dim RHO symmetries have a common ``parent'', the U(N) symmetry of the 
N-dim isotropic harmonic oscillator or the O(N+1) symmetry of the N-dim 
hydrogen atom, but they are not in general related to unitary or orthogonal
symmetries themselves.  

Since the RHO is of current interest in relation to various physical 
systems (superdeformed and hyperdeformed nuclei \cite{Mot,Rae,Ros,Bha,Naz}, 
Bloch--Brink $\alpha$-cluster
model \cite{RZ,ZR,Bri}, deformed atomic clusters \cite{Mar,Bul}), 
the unification of the rich variety 
of symmetries appearing in the RHO for different frequency ratios 
in a common algebraic framework is an interesting project. In \cite{MV}
the 3-dim RHO degeneracies  are obtained as reducible representations 
of U(3). It could be possible to construct an algebraic framework in 
which the RHO degeneracies occur as irreducible representations of an 
appropriate algebra. Work in this direction is in progress \cite{BDKL}. 

Throughout this paper the properties of the completely symmetric irreps 
of U(N) and O(N+1) have been considered. Similar studies of completely 
antisymmetric irreps, or irreps with mixed symmetry, might be worth exploring.

{\bf Acknowledgements}

One of the authors (DB) has been supported by the EU under contract 
ERBCHBGCT930467.  This work has also been supported by the Greek General 
Secretariat of Research and Technology under contract PENED95/1981.



\begin{thebibliography}{99}

\bibitem{JM}
J. M. Jauch  and E. L. Hill, {\it Phys. Rev.} {\bf 57} (1940) 641.

\bibitem{Dem}
Yu. N. Demkov, {\it Zh. Eksp. Teor. Fiz.} {\bf 44} (1963) 2007 
[{\it Soviet Phys. JETP} {\bf 17} (1963) 1349]. 

\bibitem{Dui}
F. Duimio and G. Zambotti, {\it Nuovo Cimento} {\bf 43} (1966) 1203. 

\bibitem{Mai}
G. Maiella, {\it Nuovo Cimento} {\bf 52} (1967) 1004.  

\bibitem{Ven}
I. Vendramin, {\it Nuovo Cimento} {\bf 54} (1968) 190. 

\bibitem{MV}
G. Maiella and G. Vilasi, {\it Lett. Nuovo Cimento} {\bf 1} (1969) 57.  

\bibitem{Cis}
A. Cisneros and H. V. McIntosh, {\it J. Math. Phys.} {\bf 11} (1970) 870. 

\bibitem{Mot}
B. Mottelson, {\it Nucl. Phys. A} {\bf 522} (1991) 1c.  

\bibitem{Rae}
W. D. M. Rae, {\it Int. J. Mod. Phys. A} {\bf 3} (1988) 1343. 

\bibitem{Ros}
G. Rosensteel and J. P. Draayer, {\it J. Phys. A} {\bf 22} (1989) 1323. 

\bibitem{Bha}
D. Bhaumik, A. Chatterjee and B. Dutta-Roy, {\it J. Phys. A}
{\bf 27} (1994) 1401. 

\bibitem{Naz}
W. Nazarewicz and J. Dobaczewski,  {\it Phys. Rev. Lett.} {\bf 68} (1992) 154. 

\bibitem{Nol}
P. J. Nolan and P. J. Twin, {\it Ann. Rev. Nucl. Part. Sci.} {\bf 38} (1988)
533. 

\bibitem{Jan}
R. V. F. Janssens and T. L. Khoo, {\it Ann. Rev. Nucl. Part. Sci.} {\bf 41}
(1991)  321. 

\bibitem{RZ}
W. D. M. Rae and J. Zhang, {\it  Mod. Phys. Lett. A} {\bf 9} (1994) 599. 

\bibitem{ZR}
J. Zhang, W. D. M. Rae and A. C. Merchant, {\it Nucl. Phys. A} {\bf 575} 
(1994) 61. 

\bibitem{Bri}
D. M. Brink, in {\it Proc. Int. School of Physics, Enrico Fermi Course 
XXXVI, Varenna 1966},  ed. C. Bloch (Academic Press, New York, 1966) p. 247. 

\bibitem{Mar}
T. P. Martin, T. Bergmann, H. G\"ohlich and T. Lange, {\it Z. Phys. D} {\bf 19}
(1991) 25. 

\bibitem{Bul}
A. Bulgac and C. Lewenkopf, {\it Phys. Rev. Lett.} {\bf 71} (1993) 4130. 

\bibitem{Rav}
D. G. Ravenhall, R. T. Sharp and W. J. Pardee, {\it Phys. Rev.} {\bf 164} 
(1967) 1950. 

\bibitem{BDKL}
D. Bonatsos, C. Daskaloyannis, P. Kolokotronis and D. Lenis, 
preprint hep-th/9411218.

\bibitem{Patras94}
D. Bonatsos, C. Daskaloyannis, P, Kolokotronis and D. Lenis, in {\it
Proceedings of the 5th Hellenic Symposium on Nuclear Physics (Patras 1994),
Advances in Nuclear Physics} EUR 16302, ed. C. Syros and C. Ronchi
(European Commission, Luxembourg, 1995) p. 14. 

\bibitem{Hec}
K. T. Hecht, {\it Nucl. Phys.} {\bf 63} (1965) 177. 

\bibitem{Hol}
W. J. Holman III, {\it J. Math. Phys.} {\bf 10} (1969) 1710. 

\bibitem{All}
S. P. Alliluev, {\it Zh. Eksp. Teor. Fiz.} {\bf 33} (1957) 200 [{\it 
Soviet Phys. JETP} {\bf 6} (1958) 156]. 

\bibitem{Ham}
M. Hamermesh, {\it Group Theory and its Application to Physical Problems}
(Dover, New York, 1962).

\end{thebibliography}
\end{document}